\documentclass[aps,prl,reprint,superscriptaddress,twocolumn]{revtex4-1}
\usepackage{graphicx}
\usepackage[english]{babel} %
\usepackage[usenames]{color}
\usepackage{subfigure}
\usepackage{dcolumn}
\usepackage{bm}
\usepackage{amsmath,amsfonts,amssymb,float}
\usepackage{natbib}
\usepackage{epsfig}
\usepackage{multirow}
\usepackage{textcomp} 
\usepackage{amsmath} 
\usepackage{amsthm} 
\usepackage{amssymb}	
\usepackage{graphicx} 
\makeatletter 
\newcommand{\unit}[1]{\ensuremath{\, \mathrm{#1}}}
 
\let\baraccent=\= 
\renewcommand{\=}[1]{\stackrel{#1}{=}} 

\theoremstyle{definition}

\theoremstyle{remark}

\begin{document}


\title{Controlling fast electron beam divergence using two laser pulses}

\author{R.H.H. Scott}
\email{Robbie.Scott@stfc.ac.uk}
\thanks{The authors gratefully thank the staff of the Central Laser Facility, STFC Rutherford Appleton Laboratory, the Aquitaine Regional Council and A.R. Bell for useful discussions. This investigation was undertaken as part of the HiPER preparatory project and was funded by the UK Science and Technology Facilities Council.}

\affiliation{Department of Physics, The Blackett Laboratory, Imperial College London, Prince Consort Road, London, SW7 2AZ, United Kingdom}
\affiliation{Central Laser Facility, STFC, Rutherford Appleton Laboratory, Harwell Oxford, Didcot, OX11 0QX, United Kingdom}
\author{C. Beaucourt}
\affiliation{Univ. Bordeaux/CNRS/CEA, CELIA, UMR 5107, 33405 Talence, France}
\author{H.-P. Schlenvoigt}
\affiliation{LULI, \'Ecole Polytechnique, UMR 7605, CNRS/CEA/UPMC, Route de Saclay, 91128 Palaiseau, France}
\author{K. Markey}
\affiliation{Central Laser Facility, STFC, Rutherford Appleton Laboratory, Harwell Oxford, Didcot, OX11 0QX, United Kingdom}
\author{K.L. Lancaster}
\affiliation{Central Laser Facility, STFC, Rutherford Appleton Laboratory, Harwell Oxford, Didcot, OX11 0QX, United Kingdom}
\author{C.P. Ridgers}
\affiliation{Clarendon Laboratory, University of Oxford, Parks Road, Oxford OX1 3PU, United Kingdom}
\author{C.M. Brenner}
\affiliation{SUPA, Department of Physics, University of Strathclyde, Glasgow, G4 0NG, United Kingdom}
\affiliation{Central Laser Facility, STFC, Rutherford Appleton Laboratory, Harwell Oxford, Didcot, OX11 0QX, United Kingdom}
\author{J. Pasley}
\affiliation{Department of Physics, University of York, York, YO10 5DD, United Kingdom}.
\author{R.J. Gray}
\affiliation{SUPA, Department of Physics, University of Strathclyde, Glasgow, G4 0NG, United Kingdom}
\author{I.O. Musgrave}
\author{A.P.L Robinson}
\affiliation{Central Laser Facility, STFC, Rutherford Appleton Laboratory, Harwell Oxford, Didcot, OX11 0QX, United Kingdom}
\author{K. Li}
\affiliation{GoLP, Instituto de Plasmas e Fus\~ao Nuclear - Laborat\'orio Associado, Instituto Superior T\'ecnico, 1049-001 Lisboa, Portugal}
\author{M.M. Notley}
\affiliation{Central Laser Facility, STFC, Rutherford Appleton Laboratory, Harwell Oxford, Didcot, OX11 0QX, United Kingdom}
\author{J.R. Davies}
\affiliation{GoLP, Instituto de Plasmas e Fus\~ao Nuclear - Laborat\'orio Associado, Instituto Superior T\'ecnico, 1049-001 Lisboa, Portugal}
\author{S.D. Baton}
\affiliation{LULI, \'Ecole Polytechnique, UMR 7605, CNRS/CEA/UPMC, Route de Saclay, 91128 Palaiseau, France}
\author{J.J. Santos}
\affiliation{Univ. Bordeaux/CNRS/CEA, CELIA, UMR 5107, 33405 Talence, France}
\author{J.-L. Feugeas}
\affiliation{Univ. Bordeaux/CNRS/CEA, CELIA, UMR 5107, 33405 Talence, France}
\author{Ph. Nicola\"i}
\affiliation{Univ. Bordeaux/CNRS/CEA, CELIA, UMR 5107, 33405 Talence, France}
\author{G. Malka}
\affiliation{Univ. Bordeaux/CNRS/CEA, CELIA, UMR 5107, 33405 Talence, France}
\author{V.T. Tikhonchuk}
\affiliation{Univ. Bordeaux/CNRS/CEA, CELIA, UMR 5107, 33405 Talence, France}
\author{P. McKenna}
\affiliation{SUPA, Department of Physics, University of Strathclyde, Glasgow, G4 0NG, United Kingdom}
\author{D. Neely}
\affiliation{Central Laser Facility, STFC, Rutherford Appleton Laboratory, Harwell Oxford, Didcot, OX11 0QX, United Kingdom}
\affiliation{SUPA, Department of Physics, University of Strathclyde, Glasgow, G4 0NG, United Kingdom}
\author{S.J. Rose}
\affiliation{Department of Physics, The Blackett Laboratory, Imperial College London, Prince Consort Road, London, SW7 2AZ, United Kingdom}
\author{P.A. Norreys}
\affiliation{Department of Physics, The Blackett Laboratory, Imperial College London, Prince Consort Road, London, SW7 2AZ, United Kingdom}
\affiliation{Central Laser Facility, STFC, Rutherford Appleton Laboratory, Harwell Oxford, Didcot, OX11 0QX, United Kingdom}

\date{\today}

\begin{abstract}
This paper describes the first experimental demonstration of the guiding of a relativistic electron beam in a solid target using two co-linear, relativistically intense, picosecond laser pulses. The first pulse creates a magnetic field which guides the higher current fast electron beam generated by the second pulse. The effects of intensity ratio, delay, total energy and intrinsic pre-pulse are examined. Thermal and K$_{\alpha}$ imaging showed reduced emission size, increased peak emission and increased total emission at delays of \unit{4 - 6\ ps}, an intensity ratio of \unit{10:1} (second:first) and a total energy of \unit{186\ J}. In comparison to a single, high contrast shot, the inferred fast electron divergence is reduced by 2.7 times, while the fast electron current density is increased by a factor of 1.8. The enhancements are reproduced with modelling and are shown to be due to the self-generation of magnetic fields. Such a scheme could be of considerable benefit to fast ignition inertial fusion.
\end{abstract}

\pacs{}

\maketitle


The study of fast electron transport in high density plasmas is important for numerous applications including proton and ion beam production \cite{PhysRevLett.84.670}, isochoric heating of high density matter for opacity studies \cite{Hoarty2007115}, and fast ignition inertial fusion \cite{Tabak:1994ov}.

Electron-driven fast ignition is a promising alternative route to inertial confinement fusion, albeit much less developed than the central hot spot ignition approach. The efficiency of laser energy coupling to the DT fuel is determined by the fraction of energy absorbed into the fast electrons, their temperature, divergence, and the distance from the critical surface to the compressed core \cite{Atzeni:2005gf}. The electron beam divergence, which is addressed here, can be controlled by target manufacturing techniques \cite{campbell:4169, robinson:083105} however these have a significant impact on the target complexity and cost.

This letter describes an experimental investigation of a theoretical scheme proposed by Robinson \emph{et al.} \cite{Robinson:2008aa} to reduce the fast electron divergence using two laser pulses. The first (lower intensity) pulse accelerates electrons into the target, generating an azimuthal magnetic field within the target. The second laser pulse then accelerates the main fast electron population into the target. If the pre-generated magnetic field is of sufficient magnitude and correct geometry, the divergent main electron population is deflected towards the beam axis, thereby reducing the divergence and further reinforcing the magnetic field. 

In addition to generating magnetic fields, the first pulse alters the target front surface, affecting the laser-plasma interaction of the main pulse. Particle-In-Cell (PIC) modelling shows the first laser pulse is sufficiently intense and energetic to hole-bore through the underdense plasma ablated by the pulse's leading edge, heating it to temperatures of $\sim \unit{1\ keV}$ \cite{theobald:043102,Santos:2007jh}. This will cause the front surface to expand during the delay between the pulses. 

Previous work by Yu \emph{et al} \cite{CambridgeJournals:4179292} showed (using PIC modelling) that multiple pulses can hole-bore more effectively than an equivalent single pulse in a plasma with a density twice the critical density, $n_e \simeq 2\,n_c$. The generation of magnetic fields by using two pulses in a solid target was previously attempted experimentally by Norreys \emph{et al} \cite{0029-5515-49-10-104023}. The null results were attributed to insufficient current from the first pulse and detrimental effects caused by a pre-pulse. Markey \emph{et al} \cite{PhysRevLett.105.195008} increased the efficiency of proton acceleration by using two pulses. This was attributed to the combined effects of absorption enhancement and a two stage rear surface acceleration process, yielding an optimal pulse delay of \unit{1.5\ ps}. 


The experiment reported here provides the first experimental evidence for electron beam guiding in a solid target by using two laser pulses. It was performed using the Vulcan petawatt laser at the Central Laser Facility, Rutherford Appleton Laboratory \cite{0029-5515-44-12-S15}. The 1054 nm laser pulse contained 186$\pm11$ J (except shot $t_{delay}=7$ ps (the temporal delay between the two pulses) which had $+28$ J) of energy on target, with 20$\%$ of that energy contained within a focal spot of $\unit{7\ \mu m}$ full-width-at-half-maximum (FWHM) in a duration of 1.4$\pm0.3$ ps, yielding a peak intensity of $\unit{\sim1.0\times10^{20}\ W/cm^2}$. A new picosecond OPCPA front-end \cite{Musgrave:10} gives intensity contrast $\sim$$10^{-10}$ and energy contrast  $\sim$$10^{-7}$, the low contrast front end used for selected shots had an energy contrast of $8\times10^{-5}$. The Al($\unit{75\ \mu m}$)Cu($\unit{10\ \mu m}$)Al($\unit{1\ \mu m}$) layered planar targets (transverse dimensions 5 mm) were shot (on the thicker Al layer) at 45$^{\circ}$ p-polarization. 

Two laser pulses were created by passing the incident beam through a half-wave plate and then a polarizing beam cube. The waveplate angle controls the relative pulse levels ($R$=$I_2$:$I_1$) where $I_1$ is the intensity of the first pulse on target. Roof prisms retro-reflected both pulses, the temporal delay between the two pulses ($t_{delay}$) was altered by translating one prism. The polarizations of the pulses were matched, before re-combination in a non-polarizing cube. By interfering $\unit{100\ fs}$ pulses from the seed oscillator, the pulses were synchronized to within $\unit{50\ fs}$. Calibration ensured that the sum of the energy in both pulses was constant regardless of $t_{delay}$. 

The target rear surface temperatures were measured using Cu K$_{\alpha}$ x-ray spectroscopy and streaked pyrometry of the rear surface. A KAP conical crystal with 2D spacing of 26.64 \AA \ focussed the $\unit{6.85-8.5\ keV}$  x-rays (including the Cu K$_{\alpha_1}$ and K$_{\alpha_2}$ lines) onto a FUJI BAS image plate  \cite{meadowcroft:113102}. Bulk electron temperatures within the Cu fluor layer were inferred by fitting Cu K$_{\alpha_1}$ and K$_{\alpha_2}$ line spectra generated by the non-LTE code FLYCHK \cite{Chung20053} to the data. F/5.3 optics at 59$^{\circ}$ from target normal collected the visible optical emission from the target rear surface, which was split between a spectrometer and a high speed sampling camera (HISAC)\cite{kodama:625}, the latter gave 2D spatial ($\unit{24\ \mu m}$) and temporal resolution ($\sim$$\unit{50\ ps}$) and multiframe capability (1 ns window). The Plankian thermal radiation signal was separated from the prompt optical transition radiation (OTR) signal\cite{BATON2003}, by extracting the measurement $\unit{100\ ps}$ after the laser interaction - when the OTR signal has decayed. Radiation-hydrodynamic modelling was used to back-out the initial target temperature based on the total thermal emission at $t$=100 ps. \sc{hyades} \rm \cite{PhysRevE.57.4650} was used to model the target hydrodynamic expansion and cooling, and the resultant evolution of the rear surface thermal spectrum during the 100 ps delay. The time varying thermal spectrum was folded with the spectral response of the streak camera optics, tube and spectral filtering, then spectrally integrated giving the emission intensity as a function of time for a given initial target temperature. By changing the initial target temperature, a family of intensity-time curves was generated, relating the initial target temperature to the measured intensity at $t$=100 ps. An absolutely calibrated lamp provided a reference intensity. The measured thermal emission image is caused by the fast electrons heating the target via collisional and collective mechanisms.

The fast electron spatial distribution  was measured in a $\unit{10\ \mu m}$ Cu fluor layer $\unit{1 \ \mu m}$ beneath the target rear surface. A spherically bent quartz $21\overline31$ crystal imaged the Cu K$_{\alpha}$ emission (caused by fast electron collisions) onto a FUJI BAS image plate. 

\begin{figure}[t!]
\centering
\includegraphics[scale=0.29,trim=20mm 60mm 15mm 84mm,clip]{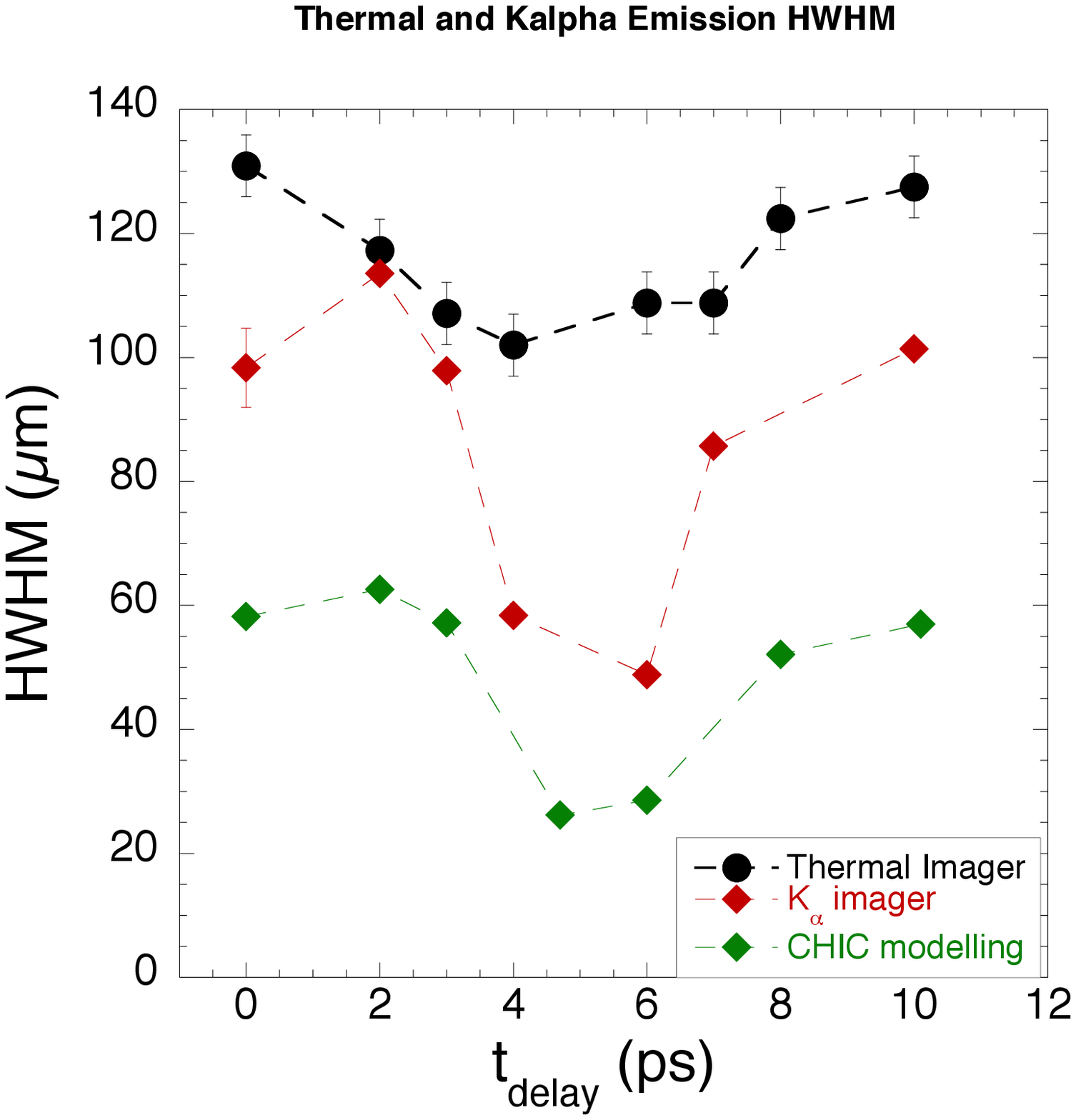}
	\caption{Dependence of the HWHM of the Cu K$_{\alpha}$ (red diamonds) and thermal emission images (dots) from the rear surface on the time delay between laser pulses. For the data at zero time delay the mean has been calculated and the standard deviation provides the error bar. The lower curve (green diamonds) shows the results of numerical simulations, this reproduces the delay required for optimal collimation.}
\label{fig:hwhm}
\end{figure}
The half-width-at-half-maximum (HWHM) of the Cu K$_{\alpha}$ and thermal emission spotsizes as a function of $t_{delay}$ are shown in figure \ref{fig:hwhm}. For the optimal $t_{delay}$=$\unit{4-6\ ps}$, the size of the K$_{\alpha}$ emission is halved, while the thermal emission is reduced by 25\%. The HWHM of the single pulse Cu K$_{\alpha}$ images increased linearly with target thickness, with a half angle of $42.0^{\circ}$ and source size of 26 \textmu m. Based on this source size, the half angle for $t_{delay}$=$\unit{6\ ps}$ was reduced to $15.4^{\circ}$ - a reduction of 2.7 times.  

The differences between the Cu K$_{\alpha}$ x-ray and thermal imaging diagnostics results depicted in figure \ref{fig:hwhm} are attributed to the thermal signal being extracted 100 ps after the initial interaction - conductivity within the target will increase the thermal emission size over this time.

\begin{figure}[t!]
\centering
\includegraphics[scale=1.0,angle=-90,trim=20mm 110mm 125mm 100mm,clip]{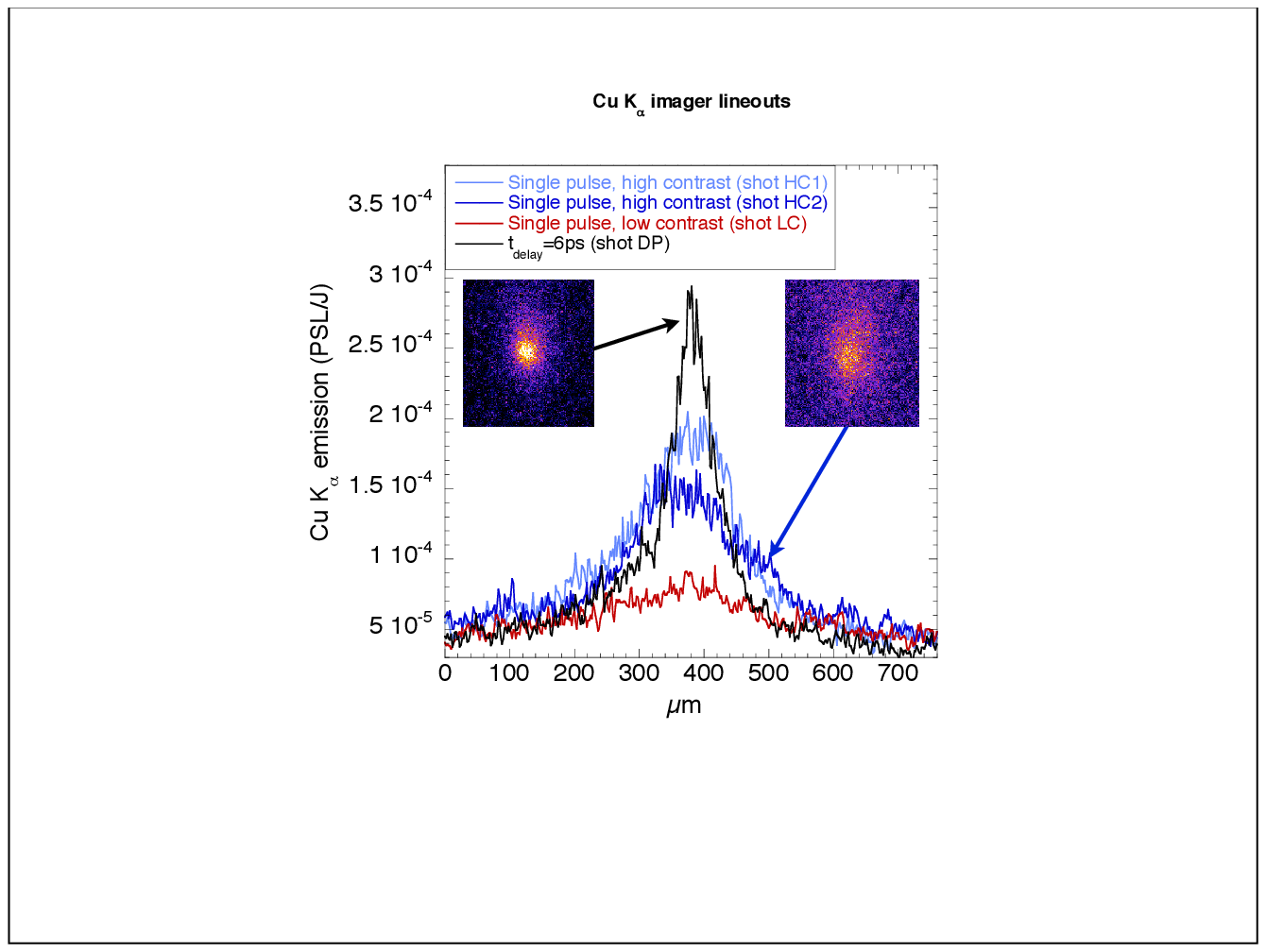}
	\caption{Examples of lineouts from the Cu K$_{\alpha}$ imager: Shots HC1 and HC2 depict the best and typical high contrast shots respectively, comparison with shot LC, which is a typical low contrast shot, illustrates the reduction in K$_{\alpha}$ spatial HWHM and increase in peak signal with increasing laser contrast. Shot DP is the optimal double pulse shot with $t_{delay}$=$6$ ps, in comparison with the high contrast data the K$_{\alpha}$ HWHM is halved while the background subtracted peak signal is increased by a factor of 1.8. Inset are two Cu K$_{\alpha}$ images with the same spatial brightness scales, showing shots HC2 and DP.} 
\label{fig:lineouts}
\end{figure}

Figure \ref{fig:lineouts} illustrates the change in the fast electron beam profile induced by the double pulse. The enhancements in the peak height and reduction in width when using the optimal parameters are clear from the double pulse shot with $t_{delay}$=$6$ps (DP), even when compared to the best single pulse shot using Vulcan's new picosecond OPCPA high contrast front-end (HC1). A comprehensive examination of the effect of the new high contrast front-end was performed by bypassing it during the experiment, reverting to the pre 2010 lower contrast \cite{Musgrave:07}. Shot LC was a typical example of Cu K$_{\alpha}$ imager data using a single, low contrast pulse; the background subtracted peak flux is enhanced by $\sim5.5 \times$ when the double pulse is used with $t_{delay}$=$\unit{6\ ps}$. It should be emphasized that this was also the first experimental implementation of the new high contrast front end on Vulcan TAP, meaning in one experiment the peak Cu K$_{\alpha}$ flux has been increased by $\sim5.5 \times$ over the previous state-of-the-art.

\begin{figure}[t!]
\subfigure[]{}\includegraphics[scale=0.24,trim=14mm 60mm 19mm 83mm,clip]{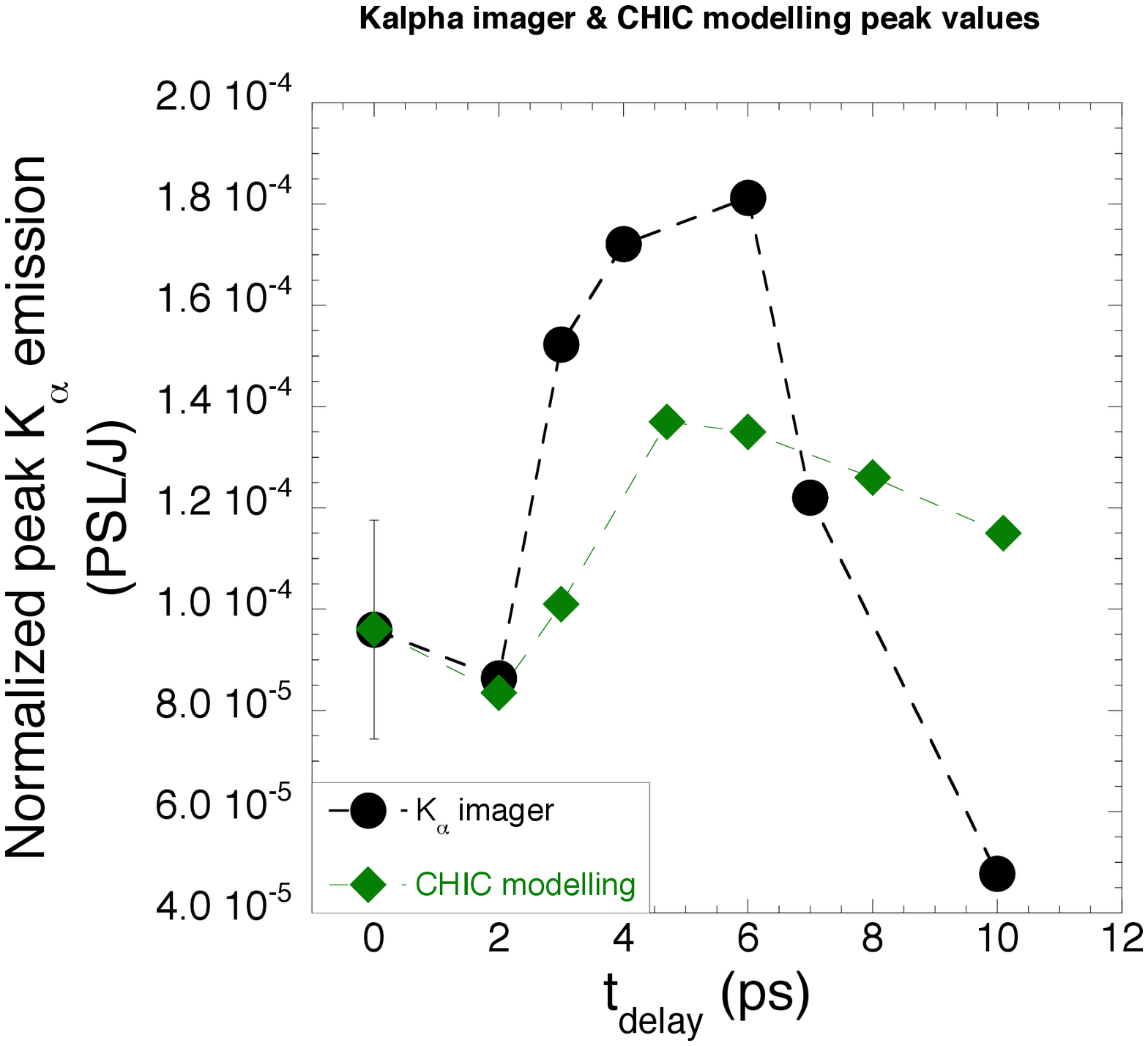}
\subfigure[]{}\includegraphics[scale=0.24,trim=14mm 60mm 25mm 83mm,clip]{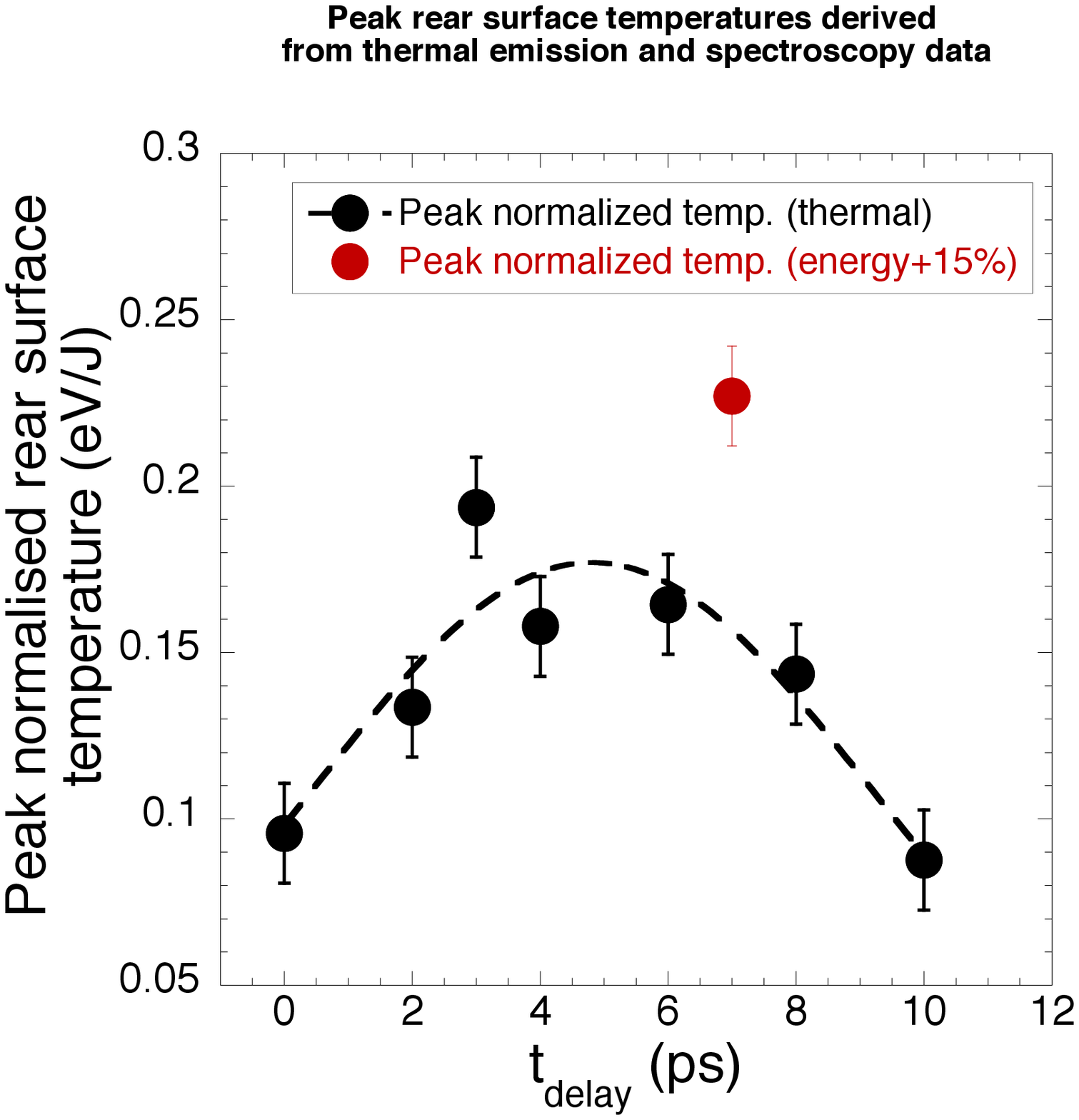}
	\caption{(a) Dependence of the peak Cu K$_{\alpha}$ emission (normalized to the laser energy on target) on the time delay obtained in the experiment (dots) and in simulations (green diamonds). The error bar at zero delay depicts the standard deviation. The modelled data is normalized to the emission at 0 ps. (b) Dependence of the peak rear surface temperatures derived from the thermal emission normalized to the laser energy on target. The data points are fitted with a Gaussian distribution. The shot in red had 15\% more energy on target than the mean, but 2.2 times higher temperature than a single pulse shot.}
\label{fig:peaks}
\end{figure}

Figure \ref{fig:peaks} (a) depicts the peak Cu K$_{\alpha}$ imager emission normalized to laser energy on target - an approximate measure of relative fast electron current density. The peak K$_{\alpha}$ emission is increased in the range $t_{delay}$ = 3-7 ps, at 6 ps the value is 1.9 times that of a single pulse. Figure \ref{fig:peaks} (b) shows laser energy normalized target rear surface temperatures. The optimal delay for both the peak Cu K$_{\alpha}$ imager emission and the thermally derived target rear surface temperatures corresponds with the optimal HWHM (figure \ref{fig:hwhm}). The mean rear surface temperature derived from Cu K$_{\alpha}$ spectroscopy was \unit{25.8\ eV} or 0.139 eV/J (standard deviation 0.028) - very similar to that of the thermal data. 

When the energy in the first pulse was halved - both by halving the total energy (90 $\pm16$ J) and keeping $R$=10:1, or by switching to $R$=20:1 (182 $\pm 20$ J) - no evidence of collimation was observed. The laser energy on the nominal shots was within 6\% of the mean except shot $t_{delay}$=$\unit{7\ ps}$ which had 15\% higher energy than the mean, interestingly the laser \emph{energy normalized} rear surface temperature (figure \ref{fig:peaks}(b)) is increased by a factor of 2.2 over the single pulse high contrast shot - considerably more than the other double pulse shots. This indicates that with more energy on target the gains due to the two pulse scheme may scale non-linearly with energy. 

The experiment was modelled using the 2D radiation hydrodynamics code \sc chic \rm \cite{Breil2007785} using MHD and fast electron transport modules. Magnetic fields are generated by the resistive electric field and the cross product of the gradients of the electron density and temperature, while the electric field is calculated assuming total current neutralization. The plasma resistivity was described by the Spitzer formula above \unit{\sim100\ eV} and by an interpolation formula \cite{PhysRevB.75.195124} below.  Fast electron transport is modelled with a reduced kinetic model  \cite{refId0,PhysRevE.84.016402} which includes self-consistent magnetic fields and collisions with plasma electrons and ions. The intensity of Cu K$_{\alpha}$ emission was calculated with a post-processor. 

Two temporally separated electron populations were injected into a 80 \textmu m thick Al target. Both electron beams have the same duration of \unit{2\ ps} FWHM. Their radial profiles at the front side were Gaussian distributions of FWHM 34 \textmu m and order 0.7 (i.e. $exp[-(r/R_0)^{(2x0.7)}]$). The energies in the first and second electron beams were \unit{1.2\ J} and \unit{15\ J}. This accounts for 20\% of the laser energy within the focal spot and absorption fractions of 33\% and 42\% for the first and second beams, respectively. The energy distribution of fast electrons were Maxwellian with temperatures of \unit{0.6\ MeV} and \unit{2.75\ MeV} calculated by taking the maximum of either Beg or ponderomotive scaling laws. The angular distribution at the source was chosen according to Ref. \cite{PhysRevE.82.036405} with a half angle divergence of 35$^\circ$ and the dispersion angle 45$^\circ$.   

As shown in figure \ref{fig:hwhm}, the modelling and experiment both have a factor of two decrease in the K$_{\alpha}$ emission HWHM with respect to the diameter from the single pulse shot, the minimum HWHM also occur with $t_{delay}$ = 4-6 ps. In both the experiment and model for $t_{delay} >$ 6 ps the HWHM increases back to the single pulse value. The minimum in the HWHM corresponds with a modelled peak magnetic field of 0.45 MG, for larger delays the magnetic field diffuses away from the propagation axis, reducing in magnitude. We note that in both the experiment and modelling, in comparison to the 0 ps case, the 2 ps delay has a slightly increased K$_{\alpha}$ HWHM. This is because in the 2ps case, $t_{delay}$ is of the order of the pulse duration, meaning the fast electron beam is effectively one longer pulse with reduced current in comparison to the 0 ps case. In this resistivity/temperature regime, the lower current of the 2 ps case generates a smaller magnetic field meaning the electron beam is less well guided. Differences in the absolute size from model to experiment may be due to differences between the modelled and experimental spatial distributions of the injected electrons. Experimentally, 20\% of the total laser energy is contained within the focal spot FWHM and 50\% is contained within a 16 \textmu m spot diameter \cite{Patel:2005aa}. Only the energy within the focal spot is modelled which may explain the differences between the experimental and modelling results. Note that reproduction of the experimental results required a modelled electron spatial distribution with `wings', this fits with the laser energy spatial distribution.

\begin{figure}[h]
\includegraphics[scale=.35,trim=20mm 41mm 0mm 42mm,clip]{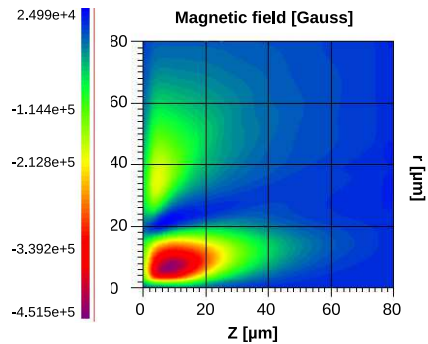}
	\caption{The modelled magnetic field at the time when the second pulse is beginning for the case where the guiding was optimal ($t_{delay}$ = 4.7 ps). The negative field near the axis collimates the fast electrons.}
\label{fig:bfield}
\end{figure}


The main features of the observed fast electron transport are explained as follows: (1) the first laser pulse `injects' a population of fast electrons into the target which seed an azimuthal magnetic field, the energy density of which is proportional to the radial derivative of the electron beam current density. On axis the induced magnetic field acts to collimate the injected fast electrons, while at larger radii the sign of the magnetic field reverses, causing the injected fast electrons to diverge. (2) During the time between the two laser pulses the induced magnetic field increases in magnitude until approximately the end of the first pulse and moves deeper into the target. Then it diffuses radially reducing in magnitude. (3) If the second electron population arrives too early or the current in the first pulse is too small, the generated magnetic field is insufficient and the beam less well guided. This explains why no collimation is observed for small $t_{delay}$, when the total energy on target is reduced and when the first pulse energy on target is reduced ($R$=20:1). (4) At the optimal delay, the magnetic field is at a maximum and the majority of the injected fast electrons interact with the convergent region of the magnetic field. In this case the optimal collimation of the beam occurs. In the case where the energy on target was higher ($t_{delay}$ = 7 ps), the current is higher meaning the magnetic field is larger and the guiding more effective. (5) The optimal delay is exceeded when the radial field diffusion is such that the collimating magnetic field is too weak to collimate the injected fast electrons, hence little guiding occurs and the beam is relatively unaffected. 


In summary, the first evidence has been provided that two laser pulses of total energy \unit{186\ J}, energy ratio 10:1 and time delay of \unit{4-6\ ps} yield optimized electron beam guiding characteristics. In comparison to single pulse shots the optimized fast electron beam has the following parameters: K$_{\alpha}$ imager HWHM $\times 0.5$, peak K$_{\alpha}$ imager signal vs high (low) contrast single pulse $\times 1.8$ ($\times 5.5$), peak thermally derived rear surface temperatures $\times 2$. Modelling accurately reproduces this data showing that a magnetic field generated within the target by the first pulse acts to collimate the second pulse. Under the optimal conditions the beam divergence is reduced by a factor of 2.7 with the fast electrons generated by the main laser pulse being guided over a distance of \unit{80\ \mu m}. This experimental evidence shows that the fast electron beam characteristics can be \emph{significantly} enhanced over the previous state-of-the-art, improving many of the fast electron beam parameters critical for fast ignition inertial confinement fusion and many other applications of intense laser-solid interactions. 


\end{document}